\documentclass[11p, twocolumn]{article}
\usepackage[utf8]{inputenc}
\usepackage[margin=2cm]{geometry} 
\usepackage{url}
\usepackage{lineno}
\usepackage{soul}
\usepackage{ragged2e, amsmath}
\usepackage{epsfig,xcolor}
\usepackage[table,xcdraw]{xcolor}
\usepackage{authblk} 
\usepackage{apacite}
\usepackage{float}

\begin{document}

    \title{\textbf{Global remote sensing reveals vegetation clustering as a physical footprint of shifting aridity trends in drylands}}

    \author[1,*]{David Pinto-Ramos}
    \author[2]{Marcel Gabriel Clerc}
    \author[3,4]{Abdelkader Makhoute}
    \author[4]{Mustapha Tlidi}

    \affil[1]{\small \textit{Center for Advanced Systems Understanding (CASUS); Helmholtz-Zentrum Dresden-Rossendorf (HZDR), D-02826 Görlitz, Germany.}}
    \affil[2]{\small \textit{Departamento de Física and Millennium Institute for Research in Optics, Facultad de Ciencias Físicas y Matemáticas, Universidad de Chile, Casilla 487-3, Santiago, Chile.}}
    \affil[3]{\small \textit{Faculté des Sciences, Université Moulay Ismail, Dynamique des Systèmes Complexes, B.P. 11201, Zitoune, Meknès, Morocco.}}
    \affil[4]{\small \textit{Faculté des Sciences, Université Libre de Bruxelles (U.L.B), CP. 231, Campus Plaine, B-1050 Bruxelles, Belgium.}}
    \affil[*]{\small \textit{Corresponding author: d.pinto-ramos@hzdr.de}}

    \date{}

\twocolumn[
  \begin{@twocolumnfalse} 
   \maketitle
    \begin{abstract}
    Due to climatic changes, excessive grazing, and deforestation, semi-arid and arid ecosystems are vulnerable to desertification and land degradation. As aridity increases, vegetation cover often self-organizes into spatial patterns before collapsing to bare soil. While recent theoretical work has established that spatially heterogeneous yet isotropic environments induce a smooth hysteresis loop---yielding either periodic (hexagonal) patterns during degradation or disordered (clustered) patterns during recovery---empirical validation of this physical footprint at a global scale has been lacking. Here, we present an extensive empirical validation using remote sensing across eight distinct global ecosystems, coupled with historical bio-climatic databases. We demonstrate that the spatial morphology of vegetation patches acts as a direct physical footprint of the ecosystem's historical aridity trend. Our results show that ecosystems experiencing increasing aridity display periodic arrays with a defined wavelength, whereas those recovering under decreasing aridity exhibit scale-free clustering. This framework provides a non-destructive, robust satellite-based indicator for diagnosing whether a dryland ecosystem is on a degradation or recovery pathway.
    \end{abstract}
    
    \vspace{0.5cm} 
  \end{@twocolumnfalse}
]

\section*{Introduction}
Semi-arid and arid landscapes are particularly vulnerable to climate change and desertification. Considering they cover up to 45 percent of the Earth's terrestrial surface, several studies have investigated how environmental changes cause uniform vegetation cover to shift into a patchy landscape \cite{lefever1997origin, klausmeier1999regular, hillerislambers2001vegetation, VonHardenberg2001, dodorico2006patterns}. It is widely accepted that symmetry-breaking instability, mediated by two opposite feedbacks (facilitation and competition) acting on different spatial scales, causes this transition even in homogeneous environments \cite{lefever1997origin}. The resulting self-organized structures, commonly called vegetation patterns, correspond to a heterogeneous distribution of biomass across space. These coherent structures are modified by precipitation gradients, suggesting that ecosystems can undergo structural changes as the factors regulating these structures vary \cite{klausmeier1999regular, deblauwe2011environmental}.

Arid ecosystems are characterized by having potential evapotranspiration (PET$_0$) exceeding the water supply provided by rainfall (P) annually. Reduced precipitation and increased evapotranspiration rates—or, equivalently, increased aridity—directly impact vegetation function \cite{berdugo2020global, hu2021aridity}. As environmental adversity increases, a homogeneous cover degrades and gives rise to gaps, then stripes (or labyrinths), and eventually patches before collapsing onto bare ground \cite{lejeune1999model, lejeune2004vegetation}. This generic sequence has been observed in modeling approaches that incorporate water transport and/or population redistribution by competition and facilitation \cite{VonHardenberg2001, rietkerk2002self, lejeune2004vegetation, dodorico2006dryland, tlidi2008}. In addition to periodic biomass distributions, localized patches of vegetation \cite{lejeune2002localized, rietkerk2004self}, localized labyrinths \cite{clerc2021localised}, arcs and spirals \cite{tlidi2018observation}, pulses \cite{ruiz2023self}, and multistable domains of oblique stripes \cite{hidalgo2024nonreciprocal} have been observed. Similarly, other mechanisms associated with the phenomenon of non-equilibrium phase separation can generate vegetation patterns in the form of non-random biphasic structures \cite{siteur2023phase, couteron2023conservative}. Nevertheless, most of these studies have assumed homogeneous and isotropic landscapes. This approximation is far from reality, and deviations from it can have severe consequences for the system's response \cite{yizhaq2017geodiversity, de2021geodiversity, pinto2022vegetation,basson2023subsurface, echeverria2023effect,pinto2025aperiodic}. 

In natural ecosystems, spatial inhomogeneities in environmental conditions are the rule rather than the exception. These spatial inhomogeneities encompass non-uniformities in the spatial distribution of precipitation, soil properties—such as spatial irregularities in topography \cite{mcgrath2012microtopography, gandhi2018topographic}, soil depth \cite{franz2011coupling}, or moisture islands \cite{rodriguez2019tree}—atmospheric factors like wind and light, or even grazing, fire episodes \cite{adler2001effect, d2007noise, tega2022spatio}, and human activities \cite{wang2023vegetation}. Consequently, the theory of vegetation patterns in heterogeneous landscapes has gained traction over the years \cite{yizhaq2014effects, yizhaq2016effects, pinto2022vegetation, echeverria2023effect}, revealing predictions of further enhanced resilience and the avoidance of catastrophic shifts, which can contrast sharply with decreased resilience when spatial anisotropies are also included \cite{pinto2023topological, pinto2026spatial}. Building upon this, \citeA{pinto2025aperiodic} recently introduced a metric that could serve as a non-invasive diagnostic tool for arid patterned ecosystems. They demonstrated that environmental heterogeneities universally induce a hysteresis loop between two populated branches: a hexagonal-like pattern under increasing aridity scenarios, and a clustered, scale-free pattern under decreasing aridity scenarios. This robust correlation between pattern morphology and environmental history presents a powerful theoretical basis for teledetection.

In this work, we shift the focus from the theoretical framework derived by \citeA{pinto2025aperiodic} to empirically assess its global applicability. We test the hypothesis that macroscopic environmental trends dictate the morphology of vegetation spots across eight distinct ecosystems worldwide. By analyzing the spatial morphology of these patches alongside high-resolution, 40-year historical aridity trends, we find that this morphological footprint successfully correlates with environmental history in seven out of the eight ecosystems studied, validating its predictive power at a global scale. Furthermore, we address the observed discrepancy in the Mozambican ecosystem, and we discuss how simplified aridity indices and theoretical representations of heterogeneity may fail to capture the complex aerodynamic and physiological realities of specific savannas, highlighting the need for data-driven modeling and accurate parameter inference in future ecohydrological assessments.

\begin{figure*}[t!]
	\centering
	\includegraphics[width=0.75\textwidth]{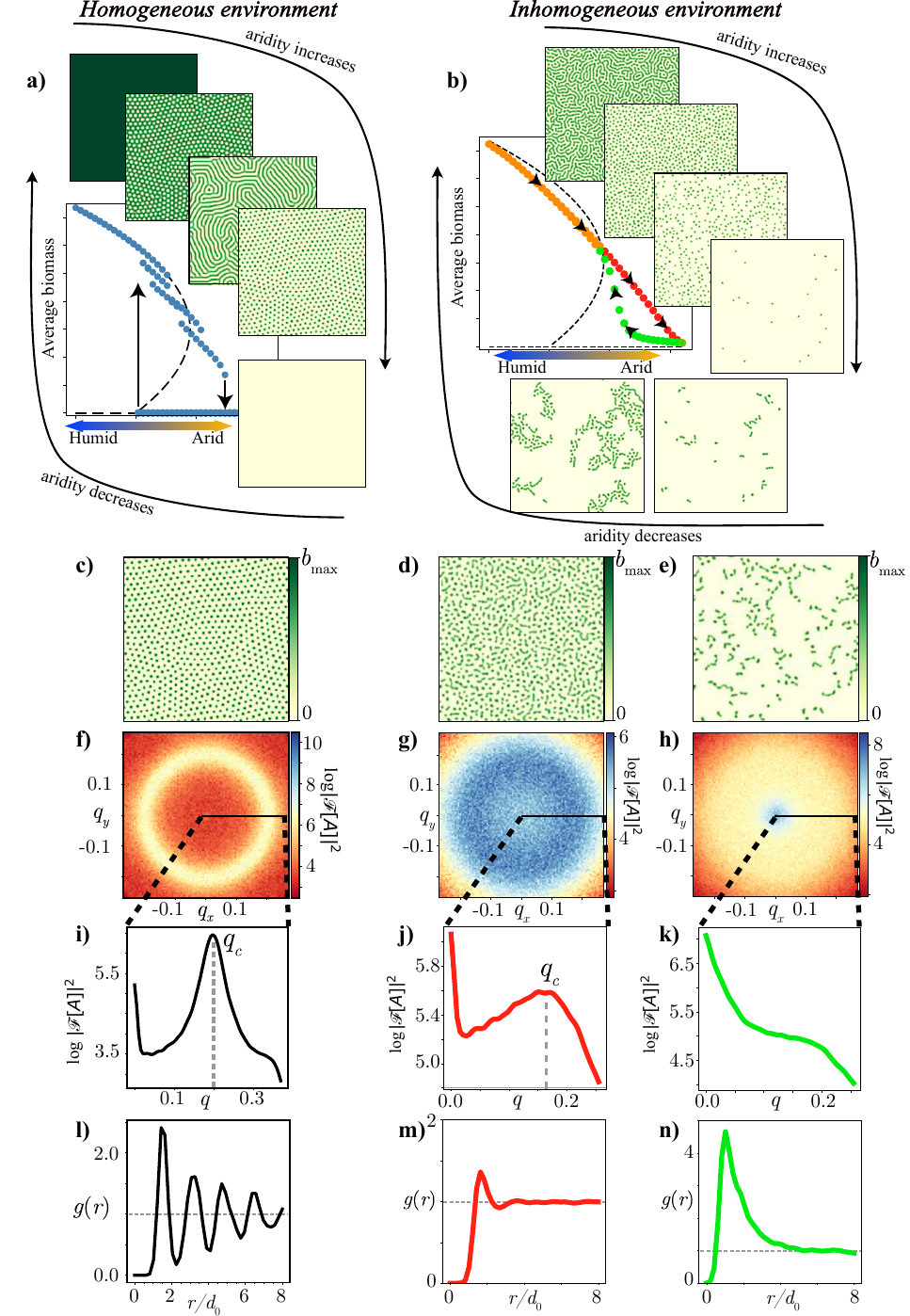}
	\caption{Conceptual representation of the hysteresis loop in pattern-forming systems under homogeneous versus heterogeneous environmental conditions, as derived in \protect\citeA{pinto2025aperiodic}. (a, b) Typical bifurcation diagrams for homogeneous and heterogeneous landscapes, with insets illustrating pattern morphology at different points within the hysteresis loops. (c) A perfect periodic spot pattern, typical of homogeneous landscapes. (d, e) Distinct spotted patterns representing coexisting equilibria for the same aridity level in a heterogeneous landscape (upper and lower branches, respectively). (f--h) Two-dimensional Fourier spectra corresponding to the patterns in c, d, and e. (i--k) Radially-averaged Fourier spectra highlighting the presence or absence of a characteristic spatial wavenumber. (l--n) Pair correlation functions of the vegetation patch positions for each respective case, distinguishing perfect crystals, disordered crystals, and scale-free clustering.}
	\label{F1}
\end{figure*}
\section*{Hysteresis loop in heterogeneous environments}
Mathematical models for vegetation spatiotemporal dynamics have established that spatial heterogeneities in environmental conditions radically alter the behavior of patterned states \cite{pinto2025aperiodic, yizhaq2014effects}. While homogeneous environments exhibit abrupt catastrophic shifts and classical sequences of regular patterns (Fig. \ref{F1}a), natural landscapes typically feature structural irregularities that break those predictions. 

As established theoretically by \citeA{pinto2025aperiodic}, introducing spatial heterogeneity into any ecohydrological model creates a fundamental hysteresis loop (Fig. \ref{F1}b). In contrast to homogeneous environments—where the hysteresis loop of equilibria occurs strictly between patterned covers and bare soil—a heterogeneous landscape induces a robust hysteresis loop between two populated branches. A spot pattern in an ideal homogeneous landscape is depicted in Fig. \ref{F1}c (the bare soil state is omitted due to its trivial spatial properties). Figures \ref{F1}d and \ref{F1}e illustrate the corresponding spotted patterns in heterogeneous landscapes for the upper and lower branches, respectively. 

Crucially, these visually similar yet intrinsically different patterns can be rigorously distinguished using their Fourier spectrum, jointly with the pair-correlation function of the spot positions. Figures \ref{F1}f--k display the Fourier spectra and their corresponding radial averages. Patterns in homogeneous landscapes showcase a perfect ring of high amplitudes, which is the classical signature of a perfect pattern (Fig. \ref{F1}f, i). On the other hand, heterogeneous landscapes exhibit powdered spectra: the upper branch retains a characteristic Fourier wavenumber indicative of a disordered crystal (Fig. \ref{F1}g, j), whereas the lower branch is entirely scale-free (Fig. \ref{F1}h, k). 

Lastly, Figs. \ref{F1}l--n present the pair correlation functions of these patch positions. The structural analysis confirms that in homogeneous landscapes, patterns behave like perfect crystals (Fig. \ref{F1}l), whereas in heterogeneous landscapes, there is a distinct bistability between a disordered hexagonal crystal (the upper branch, Fig. \ref{F1}m) and a scale-free pattern of disorganized clusters (the lower branch, Fig. \ref{F1}n). The most critical feature of this phenomenon is that these two distinct spotted morphologies are dynamically selected by aridity changes, effectively constituting a macroscopic physical footprint of environmental history \cite{pinto2025aperiodic}.

\section*{Global remote sensing image analysis}
\subsection*{Self-organization versus clustering}

\begin{figure*}[t!]
	\centering
	\includegraphics[width=0.7\textwidth]{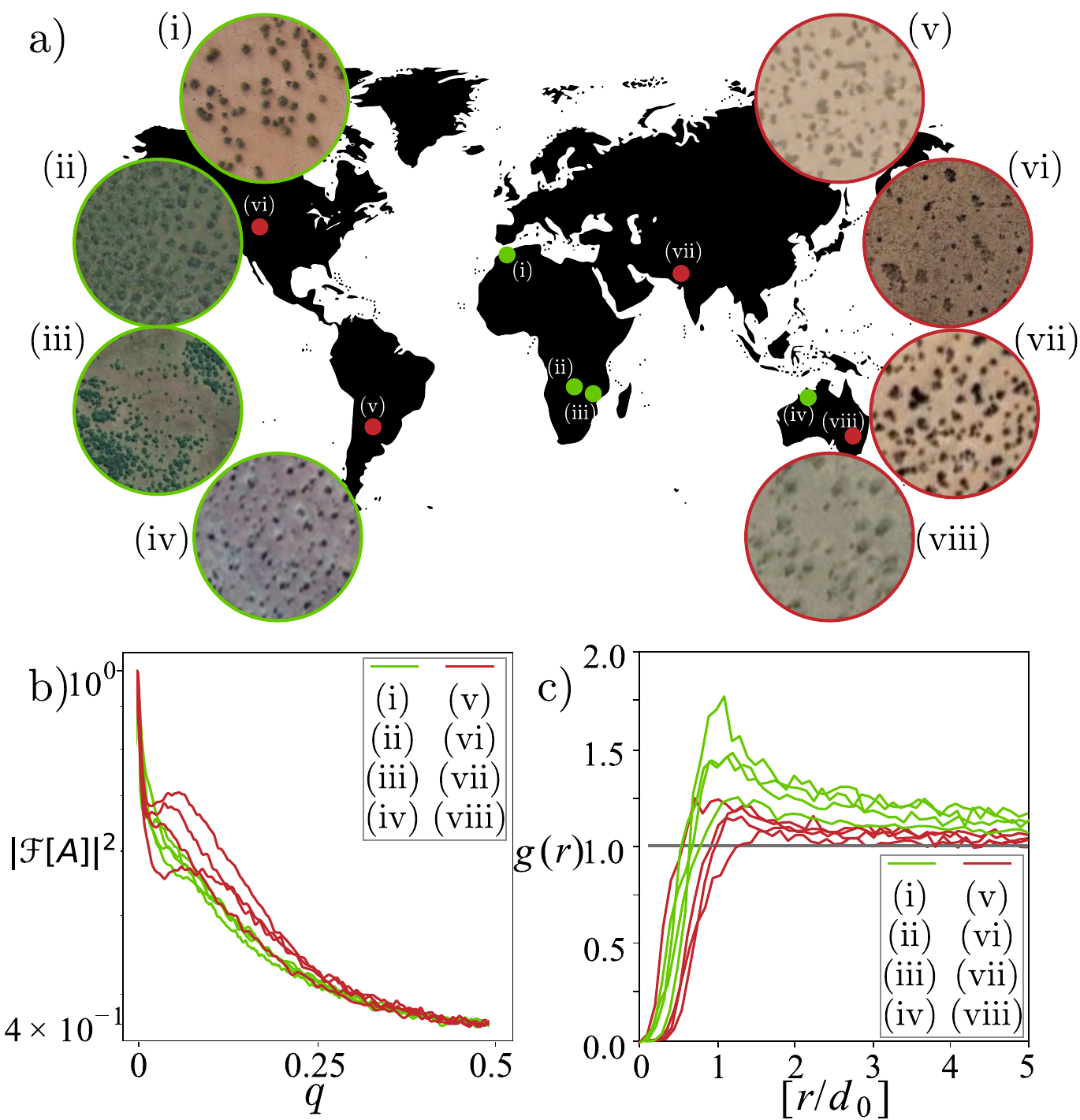}
	\caption{Vegetation patchy landscape classification across different global locations employing remote sensing analysis. 
		(a) Clustered and self-organized vegetation patches across the planet. Red and green frames [I--VIII] denote self-organized and clustered patches, respectively. The corresponding red and green dots in the map indicate the specific study locations. 
		(b) The mean radial profile of the 2D Fourier spectrum for each vegetation pattern depicted in (a). The red and green curves indicate the presence or absence of a characteristic wavelength, respectively. 
		(c) Corresponding pair correlation functions of the patch centers for each landscape, highlighting high correlation in the scale-free clustering. Satellite imagery: Google Earth Pro; map data: Google, \copyright 2022--2023 Maxar Technologies, \copyright 2023 CNES / Airbus. High-resolution UAV orthomosaic of the Moroccan ecosystem (panel I) was acquired by the authors.}
	\label{F6}
\end{figure*}

In this section, we classify the spatial statistics of spot patterns across eight ecosystems worldwide. By using spatial Fourier transforms and pair correlation functions, it is possible to rigorously categorize these spatial organizations into either periodic patterns (exhibiting a characteristic dominant wavelength) or disordered clusters (lacking a dominant wavelength). Strikingly, the presence or absence of a spatial wavelength is not specific to clonal vegetation patches observed in high-mountain ecosystems, as initially modeled by \citeA{pinto2025aperiodic}; rather, it is a universal feature found across diverse patchy landscapes with different plant functional types and soil profiles. Eight ecosystems worldwide are depicted in Fig. \ref{F6}a, where the spatial distribution of vegetation patches can be categorized based on their spectral signatures. While imagery for seven locations—spanning the United States, Argentina, Zambia, Mozambique, India, and Australia—was obtained via high-resolution satellite data (Google Earth Pro), the distinct scale of the vegetation patterns in the Moroccan ecosystem required finer spatial resolution. Consequently, high-resolution unmanned aerial vehicle (UAV) photogrammetry was deployed in Morocco to accurately capture the spatial distribution of the patches. 

The radial profile of the Fourier transform for these patchy landscapes is presented in Fig. \ref{F6}b. The vegetation ecosystems in the USA, Argentina, India, and southeast Australia demonstrate self-organization with a well-defined characteristic wavelength, in addition to the zero wavenumber (red curves, Fig. \ref{F6}b). Conversely, the green curves indicate that the vegetation patterns observed in Morocco, Zambia, Mozambique, and northwest Australia lack a characteristic wavelength; only the zero wavenumber dominates their spatial distribution. Furthermore, the pair correlation function of the patch centers confirms this structural difference (Fig. \ref{F6}c). The green curves reveal that patterns lacking a distinctive wavelength exhibit significantly higher positional correlations at shorter distances than the red curves. Interestingly, the clear exclusion zones characteristic of perfect hexagonal organization \cite{martines2013vegetation, martinez2023integrating} are not strictly observed in our empirical data. We attribute this to environmental heterogeneities being sufficiently intense to override the negative spatial correlations typically induced by resource competition.

\subsection*{The global trend map as an indicator of ecosystem self-organization and clustering}

The observation of two fundamentally different morphological states of self-organized spots raises the question of their origin. In a strictly homogeneous system, the emergence of an aperiodic distribution of spots would require extreme parameter fine-tuning or the capture of highly transient states \cite{bordeu2016self}. However, under realistic inhomogeneous environmental conditions, aperiodic distributions emerge naturally as the system dynamically responds to changing environmental stress over time. Thus, we hypothesize that the macroscopic observation of either periodic or aperiodic vegetation patches is directly coupled to the historical trajectory of local environmental conditions. 

\begin{figure*}[t!]
	\centering
	\includegraphics[width=\textwidth]{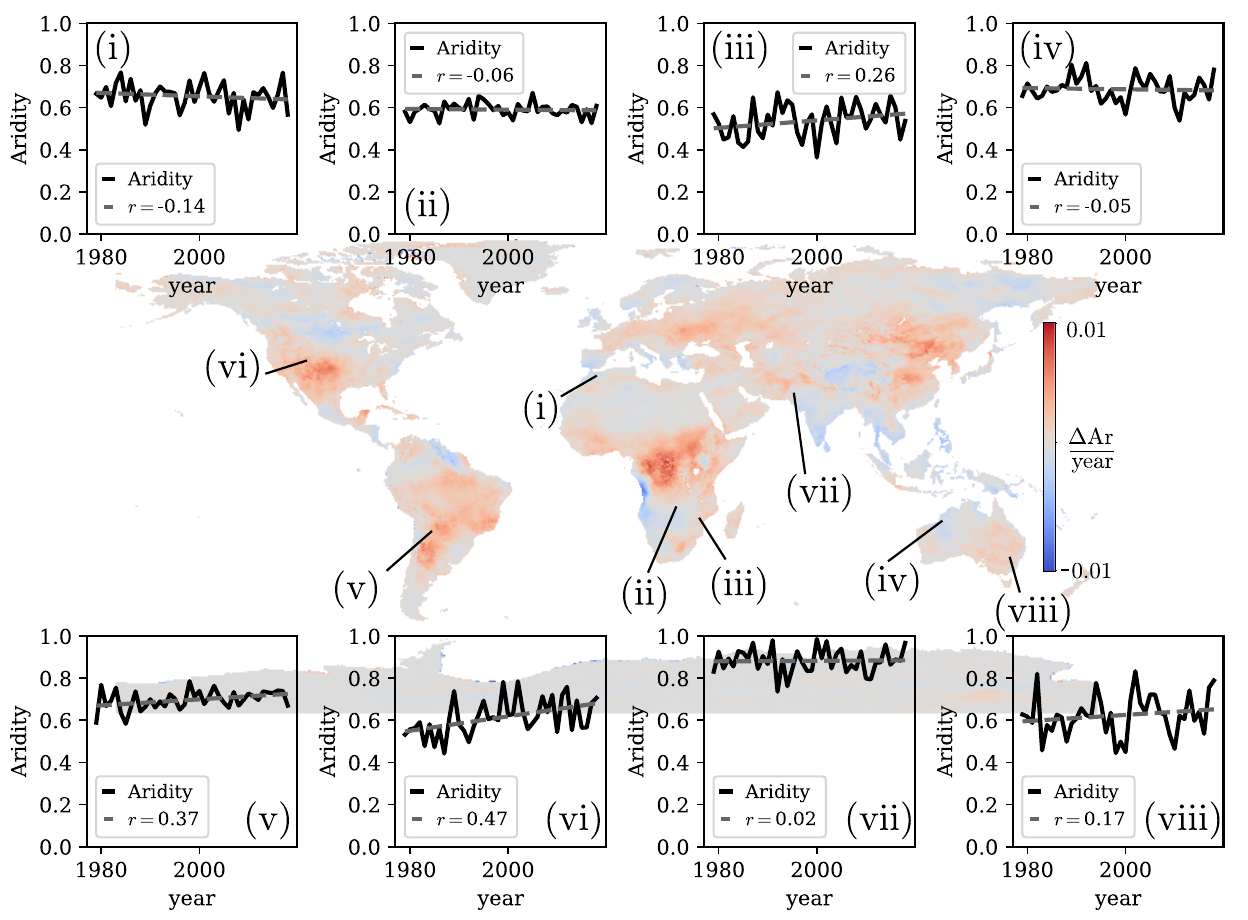}
	\caption{Global rate of change in aridity extracted from linear regression of a 40-year bio-climatic dataset (1979--2018). Blue regions indicate landscapes where aridity has decreased, whereas red regions indicate an increase in aridity. Arrows pinpoint the eight specific ecosystems analyzed. Insets display the temporal aridity trend and the corresponding Pearson correlation coefficient for each location. The observed vegetation morphologies from Morocco, the USA, Argentina, Zambia, and Australia perfectly align with the theoretical framework. The Mozambique ecosystem displays an opposing tendency, likely due to microclimatic deviations from generalized aridity indices.}
	\label{F5}
\end{figure*}

Spatial Fourier transforms and pair correlation functions successfully classify an ecosystem's structural state as either self-organized or clustered. To correlate these morphological classifications with climatic histories, we extracted a 40-year trend (1979--2018) of meteorological variables from the Copernicus Climate Change Service database \cite{copernicusData2021}. We focused on two variables crucial for patch formation and aridity assessment: mean annual precipitation (P) and reference potential evapotranspiration (PET$_0$). These variables provide global coverage at a $0.5^{\circ}$ spatial resolution. Assuming that plant mortality is a growing function of aridity, defined here as $\text{Ar} = 1 - \text{P}/\text{PET}_0$, we computed a global trend map representing the temporal gradient of environmental conditions (Fig. \ref{F5}). 

The areas shaded in red indicate a historical increase in average aridity. The ecosystems in Argentina, the USA, India, and southern Australia fall within these regions (Figs. \ref{F5} V--VIII). Crucially, all these ecosystems exhibit a self-organized spatial distribution of biomass with a well-defined dominant wavelength. Conversely, blue zones indicate regions where aridity has decreased over the 40-year period. Ecosystems in Morocco, Zambia, and northern Australia have experienced this recovering trend (Figs. \ref{F5} I, II, and V). As predicted by the heterogeneity-induced hysteresis framework, these ecosystems exhibit a clustered spatial distribution lacking a characteristic wavelength. These results suggest that the spatial self-organization of vegetation is dynamically coupled to temporal variations in aridity. As plants adapt to changing water stress, the resulting spatial footprint—either a periodic lattice with a dominant wavelength or a scale-free pattern of clusters—encodes the environmental history of the landscape. 

However, according to the global bioclimatic dataset \cite{copernicusData2021}, Mozambique's ecosystem (Fig. \ref{F5} III) does not align with this overarching trend. This discrepancy likely arises because global databases rely on simplified, large-scale estimations of potential evapotranspiration that may not fully capture the complex aerodynamic and physiological realities of specific dryland savannas. In semi-arid regions, microclimatic variables such as local wind speed, vapor pressure deficit, and the specific stomatal control of endemic plant functional types are critical components of the full Penman-Monteith equation \cite{monteith1965evaporation}. When these local factors deviate significantly from regional grid averages, simplified aridity indices can misrepresent the actual water stress experienced by the plant community, temporarily decoupling the historical PET trend from the observed spatial clustering. 

Likewise, the full complexity of natural heterogeneous landscapes is not entirely captured in the theoretical model of \citeA{pinto2025aperiodic}, nor is the detailed coupling of vegetation parameters with climate fully addressed. For example, despite the negative effects of increasing aridity on plant functioning \cite{berdugo2020global}—which drives our assumption of increased mortality rates—other bioclimatic variables, such as rising carbon dioxide concentrations, can induce opposing fertilization effects \cite{sun2022impacts}. Overall, our global macro-ecological observations provide evidence for the theoretical premise: the spatial morphology of dryland vegetation patches serves as a scale-invariant physical footprint of historical aridity trends, dictated by underlying environmental heterogeneities. Yet, our results highlight the need for more comprehensive modeling of spatial heterogeneities and plant-climate feedbacks, which may drive these complex ecosystems in intricate ways.

\begin{table*}[t!]
\centering
\caption{Morphological and Dynamical Properties of Spotted Vegetation Patterns in Inhomogeneous Environments.}
\begin{tabular}{p{0.25\textwidth} p{0.35\textwidth} p{0.35\textwidth}}
\hline
\textbf{Property} & \textbf{Disordered Clusters} & \textbf{Self-organized Hexagons} \\ 
\hline
\textbf{Fourier Spectrum} & Radial profile shows no local maxima at $q\neq0$ (scale-free). & Radial profile exhibits local maxima at the characteristic pattern wavenumber $q_c$. \\
\textbf{Pair Correlation of Positions} & High correlation up to the cluster size, followed by monotonous decay at longer distances. & Segregation at short distances. Local maxima indicate the distinct wavelength, and local minima highlight the spatial exclusion zones. \\ 
\textbf{Dynamical Origin} & Recovery pathway: Emerge from initial conditions with sparse biomass and/or a gradual reduction of environmental stress (decreasing aridity). & Degradation pathway: Emerge from initial conditions with dense biomass and/or a gradual increase in environmental stress (increasing aridity). \\ 
\hline
\end{tabular}
\label{Tab1}
\end{table*}

\section*{Conclusions}

In this study, we validated a theoretically predicted correlation between different vegetation pattern morphologies and climate trends in drylands using global remote sensing imagery. Building on the heterogeneity-induced hysteresis loop universally derived for spatially heterogeneous environments \cite{pinto2025aperiodic}, we show that the spatial morphology of vegetation patches serves as a robust physical footprint of local climatic history. By applying two-dimensional Fourier spectra and pairwise correlation functions to eight diverse ecosystems globally, we systematically classified vegetation topologies into two distinct states: periodic self-organization (disordered hexagons) and scale-free, aperiodic clustering. Notably, we observed that severe environmental heterogeneities in natural landscapes can override the expected negative spatial correlations of perfect hexagonal arrays, blurring classical exclusion zones.

By integrating these structural classifications with a 40-year global database of bio-climatic indicators (incorporating mean annual precipitation and potential evapotranspiration), we established a robust empirical correlation. Our analysis confirms that ecosystems experiencing historically increasing aridity navigate a degradation pathway, thereby forming self-organized periodic patterns with a well-defined dominant wavelength. Conversely, ecosystems situated in regions where the aridity trend has reversed navigate a recovery pathway, assembling into disordered clusters that lack a characteristic spatial scale, as summarized in Table \ref{Tab1}. While this morphological footprint proved highly consistent across most of the globe, anomalies such as the Mozambican ecosystem highlight the need to refine simple potential evapotranspiration indices using precise aerodynamic microclimate data and to account for compensatory mechanisms, such as carbon dioxide fertilization. 

Ultimately, the robust correspondence between the spatial topology of vegetation patterns and the multi-decadal history of environmental stress provides an unprecedented global validation of heterogeneity-induced hysteresis in arid landscapes. In the context of accelerating global climate change, this macro-ecological footprint offers a crucial, non-destructive early-diagnostic metric. It empowers ecologists and land managers to distinguish inherently resilient, recovering landscapes from those approaching critical degradation thresholds, relying solely on scalable satellite teledetection.
 
 \section*{Conflict of interest}
 The authors declare that they have no conflict of interest regarding this work.
 
\section*{Open Research}
Data for this study are publicly available. The bioclimatic indicators were sourced from the Copernicus Climate Change Service \cite{copernicusData2021}. Remote sensing imagery used for spatial patch analysis was derived from Google Earth Pro \cite{google_earth}. Data archiving for this imagery and the associated analysis files is currently underway. We plan to use the Rossendorf Data Repository (RODARE) for permanent hosting, and the data will be published under the DOI XXX.XXXX upon manuscript acceptance. In the meantime, a copy of the data has been temporarily uploaded as Supporting Information for peer review purposes. This temporary .zip file contains the raw images used for Fig. 2, the processed images, the exact coordinates in Google Earth as .kmz files, and the ImageJ script used to process them.

\section*{Acknowledgments}
The authors thank S. Echeverria-Alar for fruitful discussions.
M.G.C. acknowledges the financial support of ANID-Millennium Science Initiative Program-ICN17$\_$012 (MIRO) and FONDECYT project 1210353. 
M.T. is a Research Director at Fonds de la Recherche Scientifique FNRS. We would also like to thank the CNRST for its support under the FINCOME programme (N L71/2022). This work was partially funded by the Center for Advanced Systems Understanding (CASUS), which is financed by Germany's Federal Ministry of Research, Technology, and Space (BMFTR) and by the Saxon Ministry for Science, Culture and Tourism (SMWK) with tax funds on the basis of the budget approved by the Saxon State Parliament. 
\section*{Author contributions}
DPR, MGC, and MT conceptualized the work. All the authors wrote, revised, and accorded the final version of the manuscript. DPR performed the data analysis and numerical simulations. AM conducted the field campaigns and acquired the high-resolution UAV imagery for the Moroccan study site.

\bibliographystyle{apacite}
\bibliography{Vegetation_curated}

\end{document}